# Phenomenological description of bright domain walls in ferroelectric-antiferroelectric layered chalcogenides

Anna N. Morozovska[1*], Eugene A. Eliseev[2], Kyle Kelley[4], Yulian M. Vysochanskii[3], Sergei V. Kalinin[4†], and Petro Maksymovych[4‡]

[1] Institute of Physics, National Academy of Sciences of Ukraine, 46, pr. Nauky, 03028 Kyiv, Ukraine

[2] Institute for Problems of Materials Science, National Academy of Sciences of Ukraine, Krjijanovskogo 3, 03142 Kyiv, Ukraine

[3] Institute of Solid State Physics and Chemistry, Uzhgorod University, 88000 Uzhgorod, Ukraine

[4] The Center for Nanophase Materials Sciences, Oak Ridge National Laboratory, Oak Ridge, TN 37831

**Abstract**

Recently, a layered ferroelectric $CuInP_2Se_6$ was shown to exhibit domain walls with locally enhanced piezoresponse – a striking departure from the observations of nominally zero piezoresponse in most ferroelectrics. Although it was proposed that such "bright" domain walls are phase-boundaries between ferri- and antiferroelectrically ordered regions of the materials, the physical mechanisms behind the existence and response of these boundaries remain to be understood. Here, using Landau-Ginzburg-Devonshire phenomenology combined with four sub-lattices model, we describe quantitatively the bright-contrast and dark-contrast domain boundaries between the antiferroelectric, ferroelectric or ferrielectric long-range ordered phases in a layered ferroelectric-antiferroelectric ferroics, such as $CuInP_2(S_{1-y}Se_y)_6$ ($0 \leq y \leq 1$).

[*] Corresponding author, e-mail: anna.n.morozovska@gmail.com

[†] Corresponding author, e-mail: maksymovychp@ornl.gov

[‡] Corresponding author, e-mail: sergei2@ornl.gov

# I. INTRODUCTION

Multiferroics - solid state ferroic materials with coupled and ferromagnetic and antiferromagnetic, ferroelectric and antiferroelectric, or other type long-range ordering [1, 2] - have been for many years explored from a fundamental perspective [3, 4, 5], including recent studies of unusual polarization switching in thin films [6], domain wall conduction [7] and atomic scale phenomena at surfaces and interfaces [8, 9, 10, 11]. These materials are also beginning to find potential applications for applications, such as the concepts of ferroelectric tunneling barriers, light-assisted ferroic dynamics, spin-driven effects, and ultrafast magnetoelectric switching for memory [12, 13, 14].

Recently discovered multiferroics, Cu-based layered chalcogenides, with a chemical formula $CuInP_2Q_6$ ($Q$ is S or Se) [15, 16], are promising low-dimensional (e.g. single- or few-layered) uniaxial ferroelectrics [17, 18]. S- and Se-based Cu-In compounds have similar structure of individual layers, with $Cu^+$ and $In^{3+}$ ions counter-displaced within individual layers, against the backbone of $P_2Q_6$ anions [19, 20, 21]. Despite the structural similarity, the ferroelectric properties of $CuInP_2S_6$ and $CuInP_2Se_6$ are rather different [19-21]. The spontaneous polarization of the uniaxial ferrielectric $CuInP_2S_6$ ranges from 0.05 $C/m^2$ to 0.12 $C/m^2$ [22], and is about 0.025 $C/m^2$ for the uniaxial ferrielectric $CuInP_2Se_6$ [23]. The difference in the ferrielectric phase transition temperatures are ~305 K for $CuInP_2S_6$ and ~230 K for $CuInP_2Se_6$. At that $CuInP_2Se_6$ has an anomalously broad phase transition region [19-21] originating from the coexistence of ferroelectric (**FE**), or ferrielectric (**FEI**), and antiferroelectric (**AFE**) ordering, and an incommensurate phase that precedes ferroelectric ordering [20]. The properties of the intermixed S-Se compound are even more interesting [24, 25, 26], possibly involving a Lifshitz transition as well as polar glassy phases and weak dipolar correlations in the lattice [27, 28, 29]. These properties seem particularly important for prospective application of these materials as functional components of van der Waals heterostructures [30]. Indeed, recently, Song et al. [23] proposed that ultrathin films of $CuInP_2Se_6$ develop an antiferroelectric ground-state, with the crossover ferrielectric-antiferroelectric instability occurring at a thickness of ~6-8 layers. The primary driving force for the crossover is the depolarizing field that favors the antiferroelectric with net zero polarization.

An intriguing recent finding are unusual "bright" domain boundaries in $CuInP_2Se_6$, which have enhanced local piezoelectric response [31] as measured by Piezoresponse Force Microscopy (**PFM**). The effect was attributed to the coexistence of piezoelectric (FE or FEI) and non-piezoelectric (AFE) phases in $CuInP_2Se_6$, and the structure of FE-AFE domain boundaries was calculated from density-functional-theory (**DFT**). However, while the existence of these boundaries was considered plausible based on energy arguments, only qualitative agreement was obtained between observable and simulated properties (compare Figs.1-3 with Fig. 5 in Ref.[31]). Moreover, the detailed physical mechanism by which these walls become piezoelectrically active, and other relevant properties such as the emergence and mobility of these boundaries, and the applicability of these arguments to other ferroic materials have yet to be understood.

Here, using Landau-Ginzburg-Devonshire (**LGD**) approach combined with the recently developed four sub-lattices model (**FSM**) [32, 33], we explain the emergence and behavior of "bright", "mixed" or "dark" domain walls in a ferroic with coexisting AFE, FE and mixed AFE-FE long-range ordering. Our theoretical results are in a quantitative agreement with PFM results [31] obtained for Cu-based layered chalcogenide ferroelectric CuInP$_2$(S$_{1-y}$Se$_y$)$_6$, where $0 \leq y \leq 1$.

## II. LGD-FSM APPROACH

LGD-FSM hybrid approach [33] provides a link between "additional order parameters" – atomic displacements **U** of polar-active atomic groups (shown schematically in **Fig. 1**), and "intrinsic" long-range parameters, such as FE polarization **P** and AFE antipolar parameter **A**. In the framework of FSM Landau expansion of the free energy for a FE-AFE ferroic with a non-polar parent phase contains quadratic and bilinear contributions of the atomic displacements $\mathbf{U}^{(m)}$ [33] and has the form

$$G_{Landau} = \alpha_{ij}\left(\mathbf{U}^{(i)}, \mathbf{U}^{(j)}\right) + \beta_{ijkl}\left(\mathbf{U}^{(i)}, \mathbf{U}^{(j)}\right)\left(\mathbf{U}^{(k)}, \mathbf{U}^{(l)}\right) + \gamma_{ijklmn}\left(\mathbf{U}^{(i)}, \mathbf{U}^{(j)}\right)\left(\mathbf{U}^{(k)}, \mathbf{U}^{(l)}\right)\left(\mathbf{U}^{(m)}, \mathbf{U}^{(n)}\right)$$

(1)

The superscript $m$=1, 2, 3, 4 enumerates the FSM displacement vectors **U**, which corresponds to one of the four sublattices in the AFE-FE ferroic. The round brackets $\left(\mathrm{U}^{(\xi)}, \mathrm{U}^{(\zeta)}\right) = \sum_i U_i^{(\xi)} U_i^{(\zeta)}$ designate the scalar product of the corresponding vectors, where the subscript $i$=1, 2, 3 enumerates components of the vectors $U_i^{(m)}$ in the $m$-th sublattice. The derivation of Eq.(1) and link between the coefficients $\alpha_{ij}$, $\beta_{ijkl}$ and $\gamma_{ijklmn}$ with LGD-expansion coefficients can be found in Appendix A of Ref. [33].

Next, using Dzyaloshinsky substitution [34], we relate the electric polarization **P** and three antipolar order parameters (**A**, **B** and **Ã**) with the four atomic displacements $U_i^{(m)}$ of polar-active groups in a ferroic structure as [33]:

$$P_i = \frac{q}{2}\left(U_i^{(1)} + U_i^{(2)} + U_i^{(3)} + U_i^{(4)}\right), \quad A_i = \frac{q}{2}\left(U_i^{(1)} - U_i^{(2)} - U_i^{(3)} + U_i^{(4)}\right), \tag{2a}$$

$$B_i = \frac{q}{2}\left(U_i^{(1)} - U_i^{(2)} + U_i^{(3)} - U_i^{(4)}\right), \quad \tilde{A}_i = \frac{q}{2}\left(U_i^{(1)} + U_i^{(2)} - U_i^{(3)} - U_i^{(4)}\right). \tag{2b}$$

Here $q \cong \frac{Q^*}{V}$ is a dimensionality factor, proportional to the effective Born charge $Q^*$ divided by the unit cell volume $V$.

In the most common cases two combinations of atomic displacements out of four can be assumed to be zero, e.g. $\tilde{A}_i = B_i = 0$ (or $A_i = B_i = 0$). Corresponding displacements $U_i^{(m)}$ can be expressed via nonzero polar parameter $P_i$ and antipolar parameter $A_i$ (or $\tilde{A}_i$) as $U_i^{(1)} = U_i^{(4)} = \frac{P_i + A_i}{2q}$ and $U_i^{(2)} = U_i^{(3)} = \frac{P_i - A_i}{2q}$. For any case $U_i^{(1)} = U_i^{(2)} = U_i^{(3)} = U_i^{(4)} = \frac{P_i}{2q}$ in the homogeneous FE phase, while the displacements can be not equal, but of the same sign in a ferrielectric (FI) phase, which can be spatially modulated [32, 33]. The displacements $U_i^{(1)} = -U_i^{(2)} = -U_i^{(3)} = U_i^{(4)} = \frac{A_i}{2q}$, or $U_i^{(1)} = U_i^{(2)} = -U_i^{(3)} = -U_i^{(4)} = \frac{\tilde{A}_i}{2q}$, or $U_i^{(1)} = U_i^{(3)} = -U_i^{(2)} = -U_i^{(4)} = \frac{A_i}{2q}$, as well as other combinations of the alternating signs "+" and "–", corresponds to different in AFE1, AFE2, and AFE3 phases, predicted by DFT [31] (see **Fig. 1**). The case, $\tilde{A}_i = B_i = 0$, considered hereinafter, allows to make elementary algebraic transformation listed in Refs.[32, 33], and to reduce Eq.(1) to a conventional Landau formalism, which is used below. The case $A_i = B_i = 0$ can be considered in a very similar way.

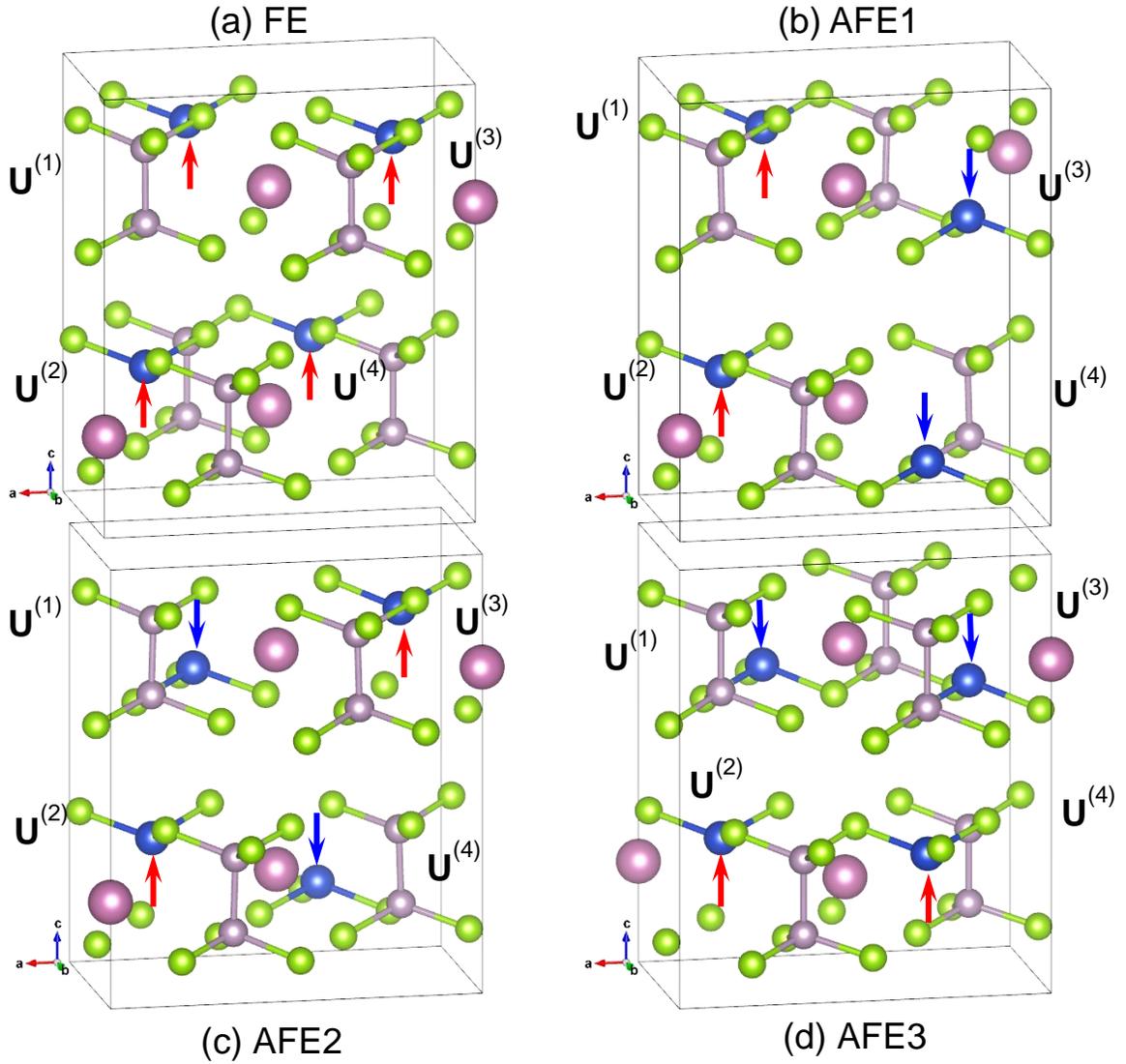

**FIGURE 1**. Atomic structures of $CuInP_2(S_{1-y}Se_y)_6$, obtained by DFT, where the small yellow and blue, small and bigger violet balls are S (or Se) and Cu atoms, P and In atoms, respectively (adapted from Ref. [31], using Open Access under a Creative Commons Attribution 4.0 International License, http://creativecommons.org/licenses/by/4.0.) Here the spatial structure is modified, and red and blue arrows inside the atomic groups are added. They illustrate the polarization direction for different types of atomic displacements $U^{(m)}$ in the quasi-homogeneous ferroelectric (or ferrielectric) FE state **(a)**, and three types of antiferroelectric states: AFE1 **(b)**, AFE2 **(c)** and AFE3 **(d)**. The orientation of the crystallographic axis a,b,c is shown at the bottom of each plot.

LGD functional of a AFE-FE ferroic utilizes Landau-type power expansion, $G_{Landau}$, that includes the contributions of FE polarization and AFE order components $P_i$ and $A_i$, as well their gradient energy $G_{grad}$, elastic and electrostriction coupling energy, $G_{els}$. For a ferroic layer confined in $x_2$-direction, the functional $G_{LGD}$ is [33]:

$$G_{LGD} = \int_{-\infty}^{\infty} dx_3 \int_{-\infty}^{\infty} dx_1 \int_0^h (G_{Landau} + G_{grad} + G_{els}) dx_2 + G_S. \qquad (3)$$

Here $h$ is the layer thickness. Electrostatic and flexoelectric contributions are neglected in this work, since we considered uncharged domain walls only, and the role of flexoelectric coupling was studied earlier [32, 33].

The Landau energy $G_{Landau}$ includes FE and AFE energies and the energy of biquadratic coupling between these order parameters:

$$G_{Landau} = G_P + G_A + G_{PA}, \qquad (4a)$$

$$G_P = a_i(T) P_i^2 + a_{ij} P_i^2 P_j^2 + a_{ijk} P_i^2 P_j^2 P_k^2, \qquad (4b)$$

$$G_A = c_i(T) A_i^2 + c_{ij} A_i^2 A_j^2 + c_{ijk} A_i^2 A_j^2 A_k^2, \qquad (4c)$$

where the summation is employed over repeated indexes. The coefficients $a_i$ and $c_i$ are temperature dependent, $a_i = a_T(T - T_C)$, $c_i = c_T(T - T_A)$, where $T_C$ and $T_A$ are the temperatures of FE and AFE phases absolute instability, respectively. Material parameters of LGD functional corresponding to CuInP$_2$S$_6$ are relatively well-known and can be found e.g. in the Table I in Ref.[35] and refs therein. LGD parameters of CuInP$_2$Se$_6$ are more uncertain (see e.g. Ref.[24-29]).

Following Kittel-type models, we assume that the temperatures $T_C$ and $T_A$ can be different and coordinate-dependent [33], at that the morphotropic phase boundary (**MPB**) between FE and AFE phases corresponds to $T_A = T_C = T_0$. Since the energy difference between AFE and FE phases are small in CuInP$_2$Se$_6$ [31], below we can consider the linear deviations of the FE and AFE temperatures from $T_0$, $T_C = T_{A0}(1 - \delta\epsilon_C(\boldsymbol{r}))$ and $T_A = T_{C0}(1 + \delta\epsilon_A(\boldsymbol{r}))$, where $|\delta\epsilon_{A,C}(\boldsymbol{r})| \ll 1$. Since the AFE transition takes place in the compound CuInP$_2$(S$_{1-y}$Se$_y$)$_6$ with Se content increase at fixed temperature and all other conditions [24-26], we can assume the equalities of temperatures $T_{A0} = T_{C0} = T_0$, and the functions $\delta\epsilon_C = \delta\epsilon_A = \delta\epsilon$; at that the assumption is in a complete agreement with Kittel-type models. The assumption allows to express "Curie-type" temperatures $T_{A,C}$ as **r**-dependent functions in terms of dimensionless parameter $\epsilon$, and correspondingly express the coefficient $a_i$ and $c_i$ through it as following [33]:

$$a_i = a_0(1 - \epsilon(\boldsymbol{r})), \qquad c_i = a_0(1 + \epsilon(\boldsymbol{r})), \qquad (5)$$

where $a_0 = a_T(T_0 - T)$ and $\epsilon(\boldsymbol{r}) = \frac{T_0}{T_0 - T} \delta\epsilon(\boldsymbol{r})$. Since the AFE transition takes place in CuInP$_2$(S$_{1-y}$Se$_y$)$_6$ with "y" increase at the fixed other conditions, we can regard the higher coefficients in Eqs.(4b)-(4c) are $\epsilon$-independent and equal, i.e. $a_{ij} = c_{ij}$, and $a_{ijk} = c_{ijk}$. Since $a_0 < 0$ at $T_0 < T$, the condition $\epsilon(\boldsymbol{r}) < 0$ supports the stability of FE phase, and the condition $\epsilon(\boldsymbol{r}) > 0$ supports the AFE phase stability. The MPB between FE and AFE phases corresponds to $\epsilon = 0$.

Note, that the parameter $\epsilon(\mathbf{r})$ plays a central role in all further theoretical analysis, and basically it is a way to control the temperatures of FE and AFE phases absolute instability, and "biasing" the system toward the desired state. In a definite sense the parameter $\epsilon(\mathbf{r})$ can be associated with the spatial changes of $T_A$ and $T_C$, being a close analogue to "random temperature" model for the Curie temperature variation in disordered (e.g. relaxor) ferroelectrics. It is worth noting that the off-stoichiometry $\delta y(\mathbf{r})$ of $y$ content in CuInP$_2$(S$_{1-y}$Se$_y$)$_6$ [24-29] can be a way to realize the parameter $\epsilon(\mathbf{r})$ in practice. Another way may be a surface chemistry.

Biquadratic coupling energy has the form [33]:

$$G_{PA} = t_{ijkl} P_i P_j A_k A_l. \quad (6a)$$

In Eq.(6a) we included only the biquadratic coupling between FE and AFE orders, assuming that the lower order coupling of $P_i$, $A_i$, and their gradients are forbidden due to the symmetry of the non-polar parent phase. The biquadratic coupling between FE and AFE orders is described by the temperature-independent tensor $t_{ijkl}$, Following Ref.[33], we regard that the strength of FE-AFE coupling is defined by dimensionless scalar parameter $\chi$:

$$t_{ijkl} = \chi t^0_{ijkl}, \quad (6b)$$

where $t^0_{ijkl}$ corresponds to the AFE material. Since the FE-AFE coupling should depend on Se content $y$, the parameter $\chi = \chi[y]$ reflecting the effect of Se-doping on the coupling strength. The FE phase can be thermodynamically stable at $0 \leq \chi \leq \chi_{cr}$ for the nonzero AFE-FE coupling, and $\chi = \chi_{cr}$ corresponds to the MPB between FE and AFE phases.

The gradient ($G_{grad}$) energy is

$$G_{grad} = g_{ijkl} \left( \frac{\partial P_i}{\partial x_k} \frac{\partial P_j}{\partial x_l} + \frac{\partial A_i}{\partial x_k} \frac{\partial A_j}{\partial x_l} \right), \quad (7a)$$

where $g_{ijkl}$ is the gradient tensor of FE and AFE long-range order parameters. Below the strengths of the gradient coefficients are introduced as

$$g_{ijkl} = g \cdot g^0_{ijkl}. \quad (7b)$$

where $g^0_{ijkl}$ corresponds to the ordered FE material, i.e. to CuInP$_2$S$_6$.

Elastic and electrostriction energy $G_{els}$ have the form:

$$G_{els} = -s_{ijkl} \sigma_{ij} \sigma_{kl} - Q_{ijkl} \sigma_{ij} (P_k P_l + A_k A_l). \quad (8)$$

Values $s_{ijkl}$ are the components of elastic compliances tensor, and $\sigma_{ij}$ is the elastic stress tensor self-consistently determined from elasticity theory equations. The last two terms in Eq.(8) are electrostriction contributions, which strength are proportional to the electrostriction tensor $Q_{ijkl}$.

The simplest form of the layer surface energy $G_S$ is conventional, $G_S = \int_{-\infty}^{\infty} dx_3 \int_{-\infty}^{\infty} \left( \frac{a_i^{(S)}}{2} P_i^2 + \frac{c_i^{(S)}}{2} A_i^2 \right) dx_1$. The non-negative values $a_i^{(S)}$ and $c_i^{(S)}$ are surface energy coefficients, which values affect the order parameter behavior near the surface of the layer. Below we consider the case of natural boundary conditions, $a_i^{(S)} = c_i^{(S)} = 0$, corresponding to $\frac{\partial}{\partial x_2} P_i = \frac{\partial}{\partial x_2} A_i = 0$.

### III. FREE ENERGY LANDSCAPE AND PHASE DIAGRAMS OF THE ORDER PARAMETERS

For a uniaxial FE-AFE ferroelectric with a second order phase transition, the free density (3) can be written in dimensionless variables as:

$$g_{LGD} = -(1-\epsilon)\frac{p^2}{2} - (1+\epsilon)\frac{a^2}{2} + \frac{p^4+a^4}{4} + \frac{\chi}{2}p^2 a^2 + \frac{g}{2}\left[\left(\frac{dp}{dx}\right)^2 + \left(\frac{da}{dx}\right)^2\right], \quad (9)$$

where the dimensionless order parameters $p = P/P_S$ and $a = A/P_S$ are introduced, where $P_S$ is a spontaneous polarization in FE phase, and we omitted all 6-order terms for a considered case of the second order phase transitions. For zero gradients, minimization of the free energy (9) allows one orthorhombic (O) and two tetragonal (T$_a$ and T$_p$) phases. The spontaneous values of the order parameter, corresponding free energy, existence and stability conditions of these phases are summarized in **Table I**, where the critical values of the FE-AFE coupling constant $\chi$, corresponding to FE-mixed phase and AFE-mixed phase boundaries, are $\chi_{cr}^p(\epsilon) = \frac{1+\epsilon}{1-\epsilon}$ and $\chi_{cr}^a(\epsilon) = \frac{1-\epsilon}{1+\epsilon}$.

**Table I.** Thermodynamic phases, order parameters, and phase boundaries

| Phase | Order parameters | Free energy | Existence | Absolute stability | Phase boundaries |
|---|---|---|---|---|---|
| T$_p$ | $p = \pm\sqrt{1-\epsilon}$ <br> $a = 0$ | $-\frac{(1-\epsilon)^2}{4}$ | $\chi > \chi_{cr}^p(\epsilon)$ and $\epsilon < 1$ | $\chi > \chi_{cr}^p(\epsilon)$ and $\epsilon < 0$ | T$_p$ with T$_a$ <br> $\epsilon = 0$ at $\chi > 1$ |
| T$_a$ | $p = 0$ <br> $a = \pm\sqrt{1+\epsilon}$ | $-\frac{(1+\epsilon)^2}{4}$ | $\chi > \chi_{cr}^a(\epsilon)$ and $\epsilon > -1$ | $\chi > \chi_{cr}^a(\epsilon)$ and $\epsilon > 0$ | T$_p$ with O <br> $\chi = \chi_{cr}^p(\epsilon)$ at $\chi^2 < 1$ |
| O | $p = \pm\sqrt{\frac{1-\epsilon-\chi(1+\epsilon)}{1-\chi^2}}$ <br> $a = \pm\sqrt{\frac{1+\epsilon-\chi(1-\epsilon)}{1-\chi^2}}$ | $-\frac{1+\epsilon^2-\chi(1-\epsilon^2)}{2(1-\chi^2)}$ | $-\frac{1-\chi}{1+\chi} < \epsilon < \frac{1-\chi}{1+\chi}$ and $\chi^2 < 1$ | $-\frac{1-\chi}{1+\chi} < \epsilon < \frac{1-\chi}{1+\chi}$ and $\chi^2 < 1$ | T$_a$ with O <br> $\chi = \chi_{cr}^a(\epsilon)$ at $\chi^2 < 1$ |

The free energy (9) as a function of polar and antipolar order parameters, $p$ and $a$, is shown in **Fig. 2** for different $\epsilon$ and $\chi$ values, and zero gradient coefficient $g = 0$.

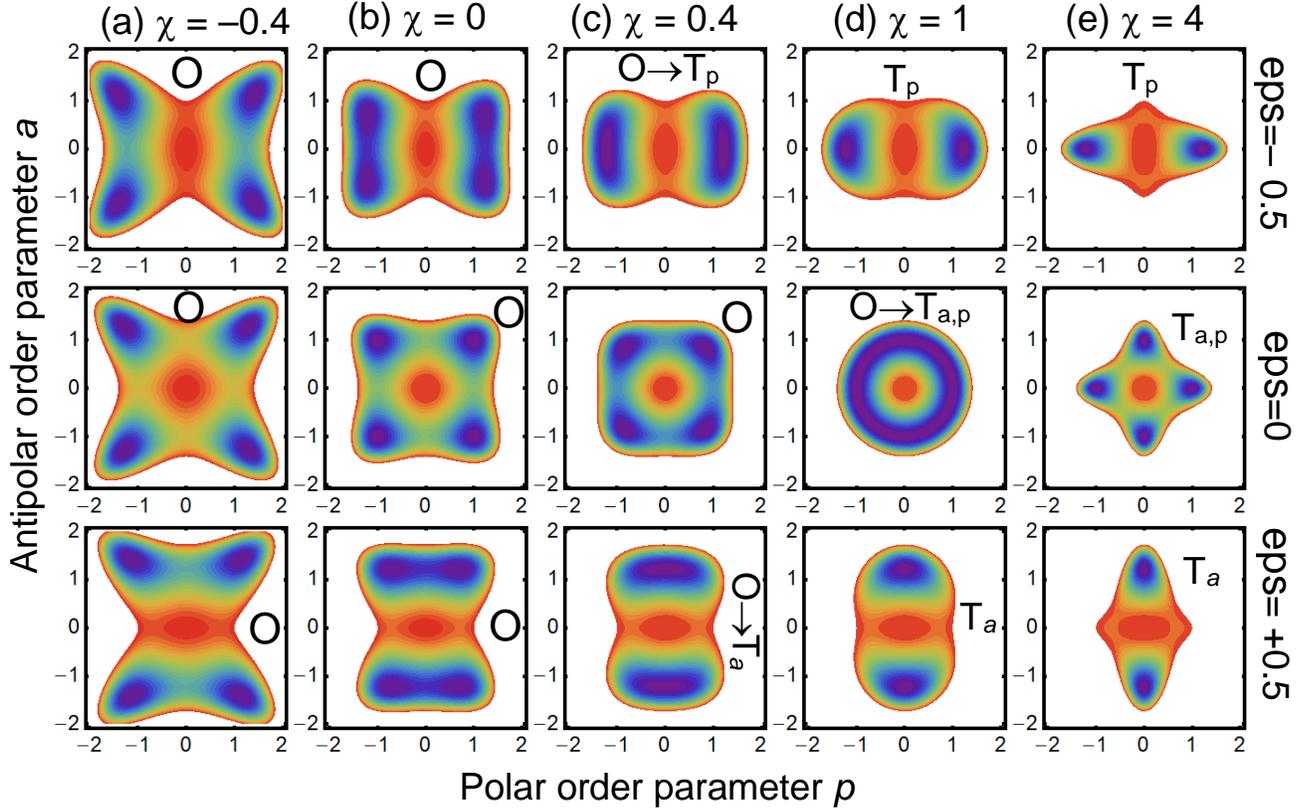

**FIGURE 2.** The free energy (8) as a function of order parameter components $p_1$ and $p_2$ for different values of the AFE-FE coupling constant $\chi$: **(a)** $\chi = -0.4$, **(b)** $\chi = 0$, **(c)** $\chi = 0.4$, **(d)** $\chi = 1$ and **(e)**. Curie temperatures change parameter $\epsilon = -0.5$ for the top line, $\epsilon = 0$ for the middle line, $\epsilon = +0.5$ and for the bottom line. Red color denotes zero energy, while violet color is its minimal density in relative units. Capital letters "O" and "$T_{a,p}$" denote orthorhombic and tetragonal spatially-homogeneous phases, respectively.

The dependencies (color map) of the order parameters $a$ and $p$ on the dimensionless transition temperatures change $\epsilon$ and FE-AFE coupling constant $\chi$, are shown in **Fig. 3a** and **3b**, respectively. The color maps of $a$ and $p$ are superimposed on the phase diagrams containing the regions of the FE, AFE, and ferrielectric AFE-FE phases. The region of the phase stability depends on the parameters $\epsilon$ and $\chi$. The quasi-homogeneous FE phase, with $p \neq 0$ and $a = 0$, is stable at $\epsilon < 0$, and the quasi-homogeneous AFE1-3 phases, with $a \neq 0$ and $p = 0$, are stable at $\epsilon > 0$. The FE-AFE coexistence boundary $\epsilon = 0$ is a straight vertical line, independent on $\chi$ and existing for $\chi > 1$. Both order parameters are nonzero in a mixed AFE-FE phase. At that, the boundary between the FE and AFE phases is $\epsilon = 0$; boundary between FE and mixed FE-AFE phases is described by equation $\chi(\epsilon) = \chi_{cr}^p(\epsilon) = \frac{1+\epsilon}{1-\epsilon}$; and the boundary boundary between the AFE and mixed FE-AFE phases is described by equation $\chi(\epsilon) = \chi_{cr}^a(\epsilon) = \frac{1-\epsilon}{1+\epsilon}$. The color maps shown in **Figs. 3a-b** are in agreement with the phase diagram shown in Fig.2b from Ref. [33].

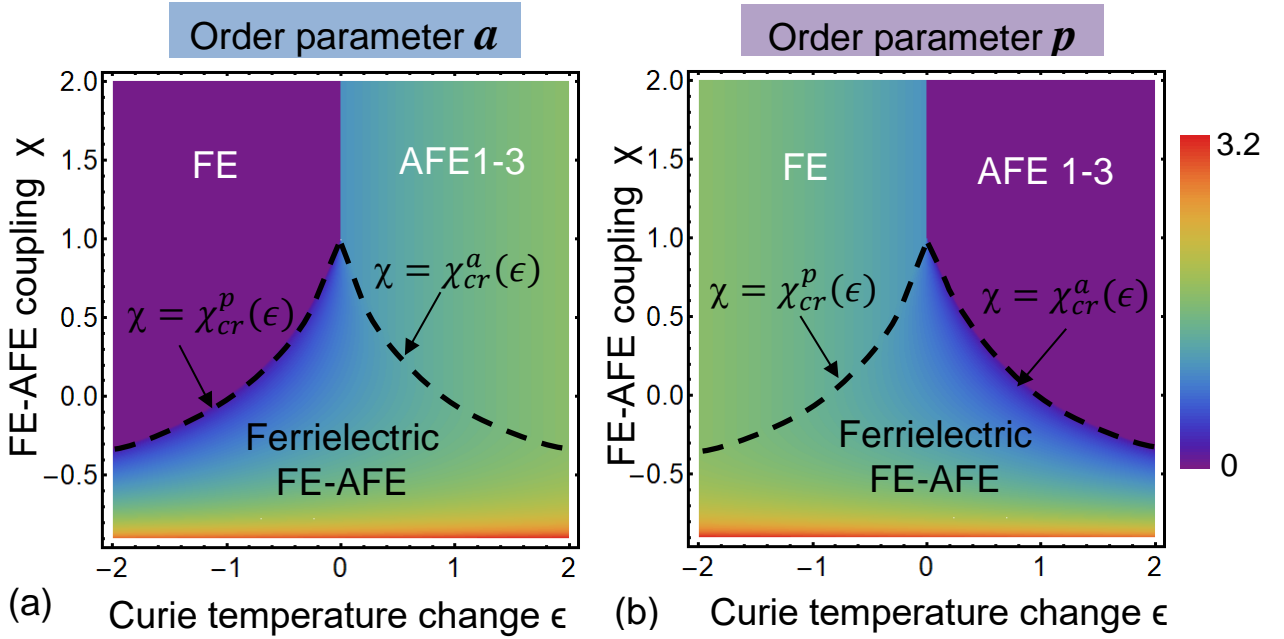

**FIGURE 3.** Color map of order parameters $a$ **(a)** and $p$ **(b)** in coordinates $\chi$ and $\epsilon$. Color bar on the right represents the color code for the both order parameters.

## IV. BRIGHT AND DARK DOMAIN WALLS

To study the coexistence of FE and AFE phases and therefore the properties of domain wall boundaries, while allowing for the gradient effects in a uniaxial ferroelectric, one should solve the coupled Euler-Lagrange equations obtained by the variation of the energy (9):

$$-[1 - \epsilon(x)]p + p^3 + \chi p a^2 - g\Delta p = 0, \qquad (10a)$$

$$-[1 + \epsilon(x)]a + a^3 + \chi p^2 a - g\Delta a = 0. \qquad (10b)$$

Here the dimensionless order parameters $p = P/P_S$ and $a = A/P_S$ are introduced, where $P_S$ is a spontaneous polarization in FE phase; and $\Delta = \frac{d^2}{dx_1^2} + \frac{d^2}{dx_2^2} + \frac{d^2}{dx_3^2}$ is a Laplacian. To avoid the emergence of depolarization field, direction **x** should be perpendicular to **P** vector. In the absence of depolarizing field, electrostatic effects do not play important role in further discussion, while in reality these effects are the main driving force for the emergence of the domain structure.

Equations (10) should be supplemented by specific boundary conditions, e.g. the order parameter periodicity (or antiperiodicity) in a homogeneous sample infinite in x-direction, e.g. $p(-L) = \pm p(L)$, $a(-L) = \pm a(L)$, where $L$ is the size of the computation region. The initial conditions can contain different numbers of uncharged $p$ and $a$ domain walls, or small random distributions of the order parameters. Note that the so-called "natural" boundary conditions, $\frac{\partial}{\partial x}p(\pm L) = \frac{\partial}{\partial x}a(\pm L) = 0$, typically leads to the system relaxation to the homogeneous state, especially in the case of initial random distributions.

Distributions of the dimensionless FE and AFE order parameters, $p$ (blue curves) and $a$ (red curves) calculated by finite element modeling (**FEM**) for coordinate-independent and coordinate-independent parameter $\epsilon(x)$ are shown in **Figs. 4a-b** and **4c-d**, respectively. FEM was performed in COMSOL@Multiphysics.

The profiles of the order parameters across the uncharged domain boundary in FE phase of the bulk material, corresponding to the negative constant $\epsilon(x) = \epsilon_0$, is shown in **Fig. 4a.** Here one can see an Ising-type 180-degree FE $p$-wall (blue tanh-like profile) with a noticeable maximum of AFE order parameter $a$ at the FE wall (red hump-like profile). The profiles of the order parameters across the domain boundary in AFE phase, corresponding to the positive constant $\epsilon(x) = \epsilon_0$, is shown in **Fig. 4b.** The picture is opposite to **Fig. 4a**, and here one can see an Ising-type 180-degree AFE $a$-wall (red tanh-like profile) with a $p$-maximum at the wall (blue hump-like profile). Hence, the is a "bright" domain wall located in the AFE phase region. We regard that the wall "brightness" is associated with polarization maximum at the domain wall.

Next, let us assume that the profile $\epsilon(x)$ has a form of a stripe of finite thickness $2x_0$, and use the "smooth" function to describe the profile, $\epsilon(x) = \epsilon_0 \left(\tanh\left(\frac{x+x_0}{L_d}\right) - 1 - \tanh\left(\frac{x-x_0}{L_d}\right)\right)$, with parameters $\epsilon_0 = 0.25$, $L_d = 10$ nm, and $x_0 = 40$ nm (**Figs. 4c**) or 80 nm (**Figs. 4d**). Typical profiles of the FE and AFE order parameters, shown in **Figs. 4c-d**, are calculated for different initial distributions of $p$ and $a$ corresponding to different number of *a*-walls in the AFE layer, and different number of *p*-walls in the FE layers near the inclusion of the AFE layer. All of these domain walls exhibit a complex structure: vanishing of one of the order parameters is accompanied by a maximum of the other one. In the cases, shown **Figs. 4c** and **4d**, there are either two "bright" domains located right outside the FE-AFE boundary, or two "bright" domain walls located inside AFE region near the AFE-FE boundary.

When a PFM scans the material in a FE phase, an image of an 180-degree domain wall is the region of gradually reduced contrast due to the decreasing FE order up to zero polarization in the center of the domain wall [31]. The structural AFE order parameter (e.g. the predicted maximum at the wall) cannot be directly observed by PFM. Thus, for a material in the AFE phase, the PFM imaging of the AFE domain walls looks opposite to the walls in the FE phase. Bulk AFE domains do not give any PFM response, the contrast appears only near the AFE antiphase domain walls with local polarization. A maximally localized PFM response will be observed here. Therefore, we can conclude that "dark" domain walls are observed by PFM in FE phase, and "bright" PFM domain walls should be observed in AFE phase, or at the AFE-FE boundaries. The conclusion is in a qualitative agreement with recent PFM experiments [31].

However, one of the features we should emphasize, is that the domain walls in our model acts as "pair-entities" caused by the symmetry of periodic boundary conditions. There is always a bright-dark pair (see **Figs. 4c-d**). In pure phases, pure FE or pure AFE, the pair is co-located, and so we see FE or AFE wall, respectively (see **Figs. 4a-b**). The pair splits in space near symmetric AFE-FE boundaries separating AFE region from FE regions, but it is still stable. The question about the pair experimental observation is discussed in next section.

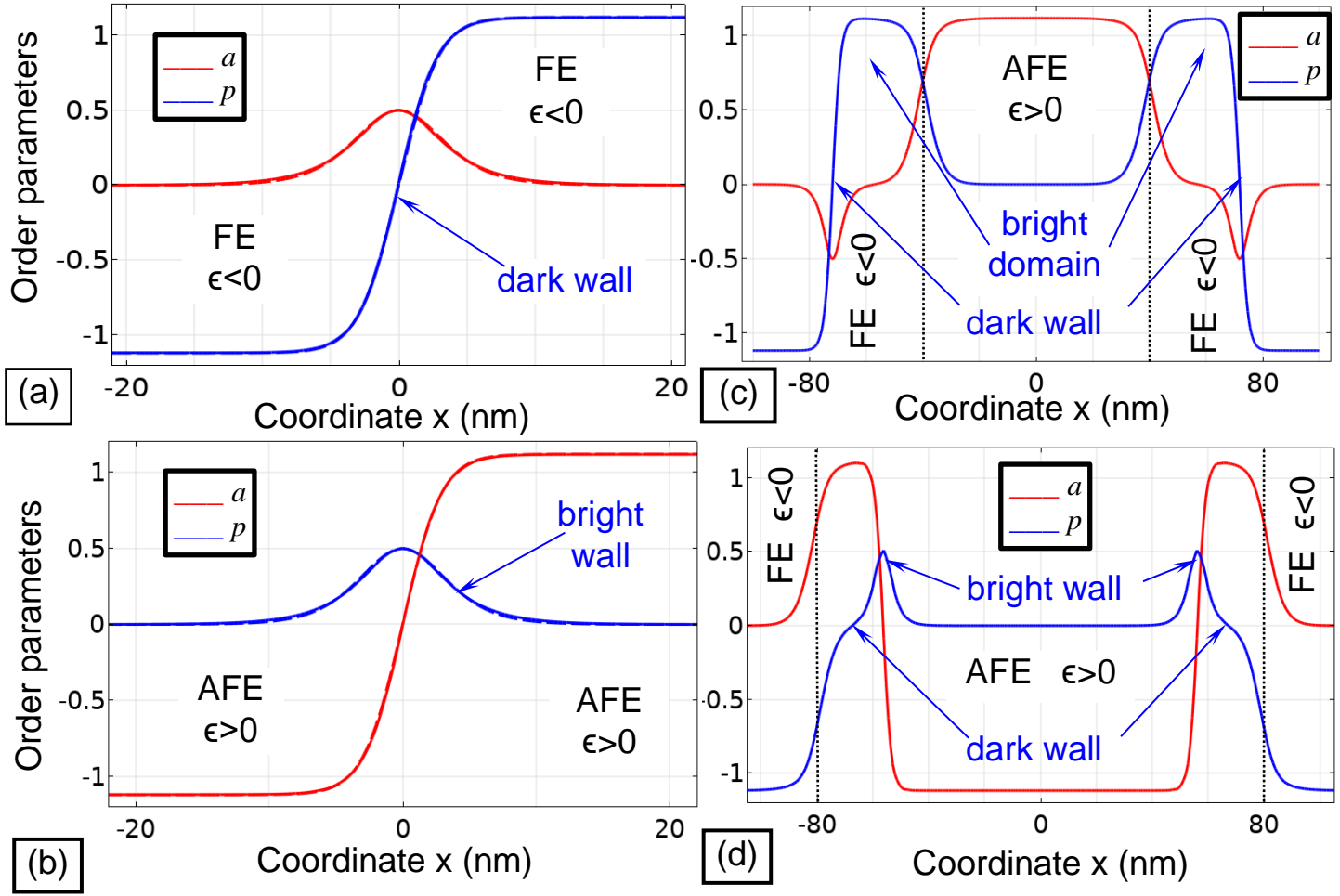

**FIGURE 4.** Distributions of the FE and AFE order parameters, $p$ (blue curves) and $a$ (red curves), respectively. Plots (a) and (b) illustrate the profiles of $a$ and $p$ at the domain wall in FE and AFE phases, corresponding to $\epsilon_0 = -\frac{1}{4}$ and $\epsilon_0 = +\frac{1}{4}$, respectively. Plots (c)-(d) show the profiles of $a$ and $p$ at the boundaries between FE and AFE phases,

induced by the function $\epsilon(x) = \epsilon_0 \left[\tanh\left(\frac{x+x_0}{L_d}\right) - 1 - \tanh\left(\frac{x-x_0}{L_d}\right)\right]$. The difference between **(c)** and **(d)** is in the initial distribution of the order parameters and AFE region width. Other parameters: $\epsilon_0 = 0.25$, $L_d = 10$ nm, $x_0 = 40$ nm **(c)** or 80 nm **(d)**, $\chi = 1$, $g = 4 \cdot$nm². Solid curves in all plots are calculated by FEM. Dashed curves in plots (a) and (d) are calculated from Eq.(11a) and (11b) with fitting parameters listed in the text. Dotted vertical lines in plots **(c)**-**(d)** denote the points where $\epsilon \approx 0$.

To analyze FEM results, shown in **Figs. 4**, we used the approximate analytical expressions for the order parameter profiles, which can be found for several specific cases, namely for $\chi \approx 0$ and arbitrary $|\epsilon(x)| < 1$, or for $\chi \approx 6[1 \pm \epsilon(x)]$ and very small $|\epsilon(x)| \ll 1$. In the first case the solution is a single tanh-profile [36], and a superposition of several tanh-profiles [37, 38] in the second case. While the tanh-profiles are not general, FEM confirms that slightly more complex trial functions can be used to describe the dark (**DW**) and bright (**BW**) p-walls in FE and AFE phases, namely:

DW: $p(x) = \frac{p_s}{2}\left[\tanh\left(\frac{x+x_w}{L_p}\right) + \tanh\left(\frac{x-x_w}{L_p}\right)\right]$, $a(x) = \frac{a_s}{2}\left[\tanh\left(\frac{x+x_w}{L_a}\right) - \tanh\left(\frac{x-x_w}{L_a}\right)\right]$, (11a)

BW: $p(x) = \frac{p_s}{2}\left[\tanh\left(\frac{x+x_w}{L_p}\right) - \tanh\left(\frac{x-x_w}{L_p}\right)\right]$, $a(x) = \frac{a_s}{2}\left[\tanh\left(\frac{x+x_w}{L_a}\right) + \tanh\left(\frac{x-x_w}{L_a}\right)\right]$. (11b)

Here the amplitudes $p_s$ and $a_s$ define FE and AFE order parameters far from the wall, because $p(x \to \pm\infty) \to \pm p_s$ for dark walls and $a(x \to \pm\infty) \to a_s$ for bright ones. The correlation lengths $L_p$ and $L_a$, and the shift $x_w$ define the width of the $p(x)$ and $a(x)$ domain walls, respectively. Also $|x| - x_w$ is the distance from center of the FE-AFE boundary $x = 0$. The height of the p-maximum located at the bright p-wall is equal to $p_s \tanh\left(\frac{x_w}{L_p}\right)$ and its width is an order of $2L_p$. Quite symmetrically, the height of the a-maximum located at the dark p-wall is equal to $a_s \tanh\left(\frac{x_w}{L_a}\right)$ and its width is an order of $2L_a$.

As an example, dashed curves in **Fig. 4a**, which are calculated from analytical expressions (11a) with parameters $p_s = 1.12$, $a_s = 2.16$, $L_p = 2.6$ nm, $L_a = 4.25$ nm, and $x_w = 1$ nm, fit with very high accuracy (point-to-point) solid curves calculated by FEM. The dashed curves in **Fig. 4b** are calculated from expressions (11b) with parameters $a_s = 1.12$, $p_s = 2.16$, $L_a = 2.6$ nm, $L_p = 4.25$ nm, and $x_w = 1$ nm, which are related with the parameters of the dashed curves in **Fig. 4a** by expected substitution $p \leftrightarrow a$.

In general, five values $p_s, a_s, L_p, L_a$ and $x_w$ are variational parameters, which can be determined after substitution of Eqs.(11) in the free energy (3), and further integration and minimization over these parameters. It appeared that the proportionalities for amplitudes $p_s \sim \sqrt{1-\epsilon}$ and $a_s \sim \sqrt{1+\epsilon}$, and for correlation lengths $L_p \sim \sqrt{\frac{2g}{1-\epsilon}}$ and $L_a \sim \sqrt{\frac{2g}{1+\epsilon}}$, are approximately valid.

The changes of local strains induced by the variation of $\epsilon$ can be estimated from expression:

$$\delta u_{in} = V_{in}\epsilon(y,\boldsymbol{x}) + d_{ink}P_k + Q^P_{inkj}(P_k P_j - P_{Sk}P_{Sj}) + Q^A_{inkj}(A_k A_j - A_{Sk}A_{Sj}), \quad (12a)$$

where $V_{in}$ is a Vegard stress tensor, $d_{ink}$ is a piezoelectric tensor, $Q^P_{inkj}$ and $Q^A_{inkj}$ are electrostriction tensors for FE and AFE order parameters, respectively. For the considered one-dimensional and one-component approximation Eq.(12a) yields to the estimate:

$$\delta u = V_g \epsilon(\boldsymbol{x}) + d\, p + Q^P(p^2 - p_S^2) + Q^A(a^2 - a_S^2). \quad (12b)$$

The main result of this section is the explanation of the domain wall structure in different AFE and FE phases of a ferroic. In particular, we expect the emergence of "dark" or "bright" domain walls at the boundary between FE and AFE phases. In accordance with our model, the spatial gradients of the FE and AFE transition temperatures, $\frac{d\epsilon}{dx}$, can lead to the coexistence of the FE, FEI, AFE, mixed FE-AFE and FEI-AFE phases, as well as to the suppressed or enhanced local electromechanical response at the boundaries between the phases. Moreover, mixed phases can be spatially modulated and at the same time incommensurate.

## V. COMPARISON WITH EXPERIMENT

Using low temperature PFM, the coexistence of piezoelectric and non-piezoelectric phases separated by unusual "bright" domain walls with enhanced piezoelectric response have been revealed in CuInP$_2$Se$_6$ [31], and explained by enhanced piezoresponse at the FE(FEI)-AFE phase boundary (see **Fig. 5a**). The AFE state was partially polarized, with inclusions of structurally different FEI domains enclosed by the "enhanced" phase boundaries, which indicates the coexistence of AFE, FEI, and FE-AFE phases, and the conclusion was supported by optical spectroscopies and DFT calculations as detailed in reference [31]. The layered ferroelectric CuInP$_2$S$_6$ only revealed "dark" domain by comparison (see **Fig. 5b**).

The "bright" and "dark" domain walls shown in **Fig. 5a** and **5b**, respectively, and are in a qualitative agreement with our theoretical results shown in **Fig. 4**. Blue and red curves in **Fig. 5a** and **5b** demonstrate the quantitative agreement between experimental data points [31] and theoretical model evolved here.

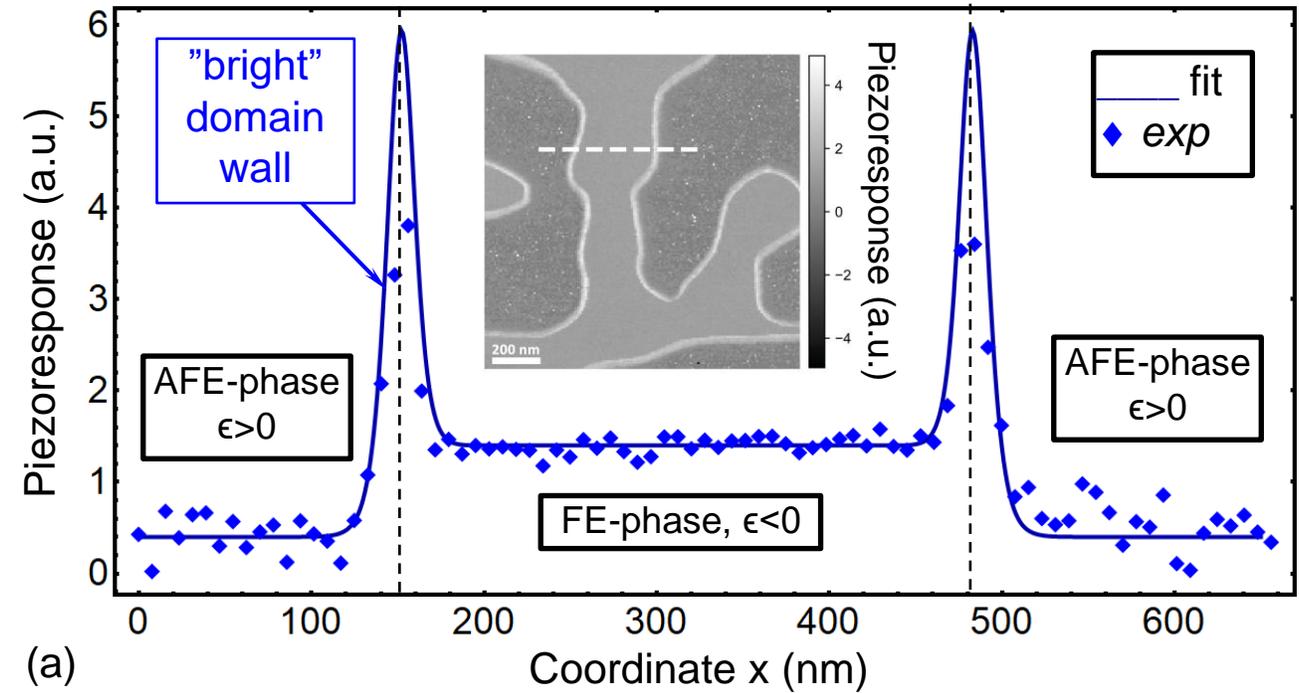

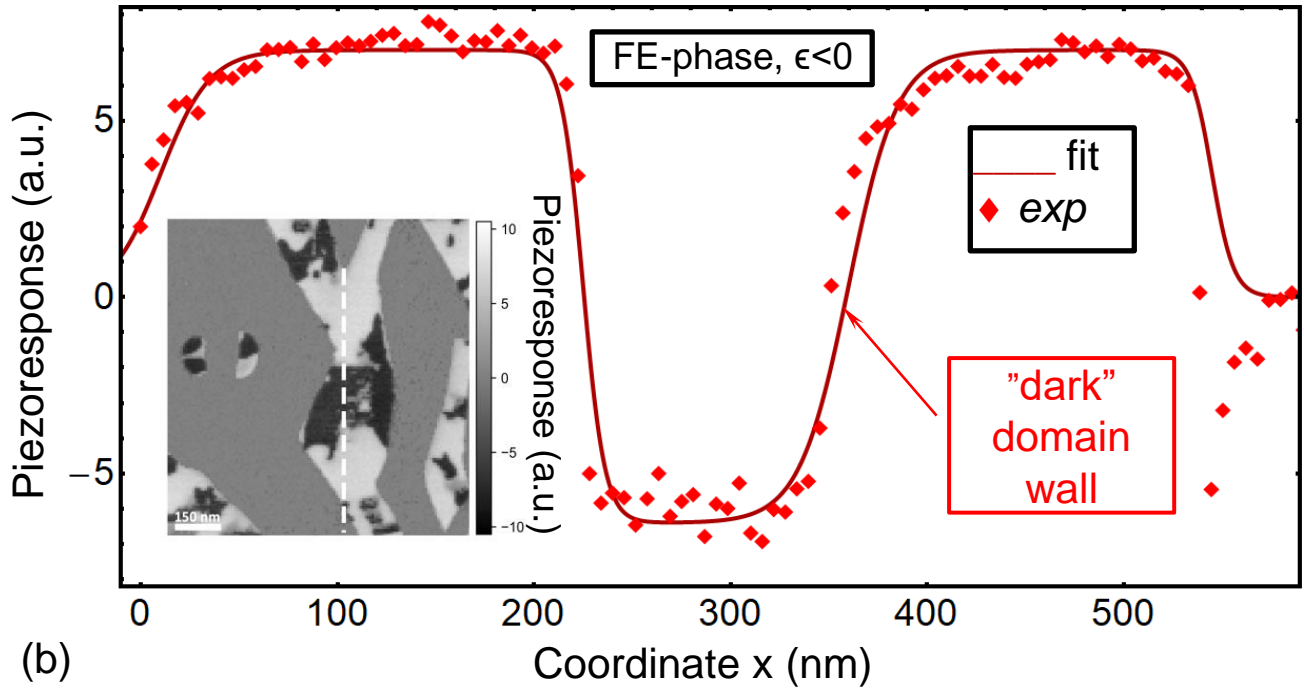

**Figure 5.** Local piezoresponse (in arbitrary units) at CuInP$_2$Se$_6$ **(a)** and Cu$_{1-x}$InP$_2$S$_6$ **(b)** surfaces. Red and blue diamonds are PFM data [31] measured at 140 K in ultrahigh vacuum for CuInP$_2$Se$_6$ and that of CuInP$_2$S$_6$ at room temperature in controlled environment. Solid dark red and blue curves are fitting by Eq.(13a) with parameters provided in the text. Black arrows point to the regions of the "bright" (a) and "dark" (b) domain walls. Line-profiles of piezoresponses (a,b) were measured along the white dashed lines, crossing domain boundaries in inset to the plots (a)

and (b), respectively. Color insets are adapted from [31], using Open Access under a Creative Commons Attribution 4.0 International License, http://creativecommons.org/licenses/by/4.0.

However, it is important that we see a bright wall in **Fig. 5a**, whereas according to **Figs. 4c-d**, it should be a bright-dark pair. We can argue that the "dark" FE wall can be anywhere in the signal, and so its detection requires careful adjustment of experimental offsets. A follow-on point to this is which of the **Fig. 4c** or **4d** is better consistent with experiment – a bright domain or a bright domain wall? The width of bright regions extracted from experimental data can help to answer. So, for detailed experimental matching, the analysis of PFM signal as a function of applied electric field (to rule out electrostatic effects) is highly recommended. Another experimental idea would be to try to switch FE and AFE domains separately from their domain boundaries. Observation of bright and dark domain walls independently from the properties of the AFE/FE boundary would serve well to understand the imaging and piezoelectric property of the boundary itself.

To fit the PFM response (**PR**) of CuInP$_2$Se$_6$, we use the kink-type profiles similar to Eqs.(11), which are inherent to the diffuse Bloch-Ising type domain walls and typical for multiaxial ferroelectrics in mixed phases [36, 38]. The functions are:

$$PR(x) = u_0 + \sum_{i=1}^{4} u_i \tanh\left(\frac{x-x_i}{w_i}\right). \tag{13a}$$

The number "4" in the sum originates from 2 bright well-separated domain walls, each of which is described by 2 tanh-functions with their own amplitudes $u_i$, intrinsic width $w_i$ and shifts $x_i$. The constant offset level is $U_0$. The best fitting to the blue symbols [31] corresponds to the following parameters $u_0 = 0.4$, $u_1 = -u_4 = 11$, $u_2 = -u_3 = -10.5$ (a.u.), $w_1 = w_4 = 11$, $w_2 = w_3 = 10$ (nm) and $x_1 = 150, x_2 = 155, x_3 = 480, x_4 = 485$ (nm). Using the symmetry of the fitting parameters the solution can be rewritten as:

$$PR(x) = u_0 + u_1\left[\tanh\left(\frac{x-x_1}{w_1}\right) - \tanh\left(\frac{x-x_4}{w_1}\right)\right] + u_2\left[\tanh\left(\frac{x-x_2}{w_2}\right) - \tanh\left(\frac{x-x_3}{w_2}\right)\right]. \tag{13b}$$

Actually, it was shown [38], that the difference (or sum) of 2 tanh-functions with the same scale parameter $w$ can be considered as trial functions describing Bloch-Ising type domain walls.

To fit the PFM response of CuInP$_2$S$_6$ in the ferroelectric phase, at first we used the Jacobi elliptic functions – "snoids", each of which are exact solutions for uncharged domain walls satisfying the static LGD equations with cubic nonlinearity and without depolarization fields [38]:

$$PR(x) = u_0 + \sum_{i=1}^{2} u_i(x)\, sn\left(\frac{x-x_i}{w_i\sqrt{1+m}}\bigg|\, m\right). \tag{13c}$$

where the constant offset $u_0$, slow-varying (due to the presence of surface defects) amplitudes $u_i(x)$, "module" $0 \leq m \leq 1$, and "shifts" $x_i$ of snoids are fitting parameters. The best fitting was to the red symbols [31] obtained using Eqs.(13c) corresponds to $m = 0.999$, i.e. it tends unity. This result means that the wall profile is strongly nonlinear, and the limit $sn\left(\frac{x-x_i}{w_i\sqrt{1+m}}\Big| m\right)\Big|_{m\to 1} \to \tanh\left(\frac{x-x_i}{2w_i}\right)$ is well-grounded. That is why, as a next stage, we used the functional form (13a) for the fitting of local PR and determine that the following best fitting parameters $u_0 = 0$, $u_1 = -u_4 = 3.5$, $u_2 = -u_3 = -6.7$ (a.u.), $w_1 = w_3 = 25$, $w_2 = w_4 = 11$ (nm) and $x_1 = 10, x_2 = 225$, $x_3 = 360, x_4 = 545$ (nm), correspond to CuInP$_2$S$_6$.

Since in both cases Eq.(13b) appeared to be an optimal fitting, we can comment on the fitting parameters and try to extract some information about the internal parameters $\epsilon(x)$, $\chi$, $g$, and $x_{wi}$ in Eqs.(11) from them. The main differences appear to be 3 times larger $u_1 = -u_4 = 11$ for CuInP$_2$Se$_6$ in comparison to $u_1 = -u_4 = 3.5$ for CuInP$_2$S$_6$; at that $u_2 = -u_3 = -10.5$ for CuInP$_2$Se$_6$ are relatively close to $u_2 = -u_3 = -6.7$ for CuInP$_2$S$_6$. We note the twice difference in the widths $w_1 = w_3 = 25$ nm and $w_2 = w_4 = 11$ nm for the CuInP$_2$Se$_6$, in contrast to almost the same widths $w_1 = w_4 = 11$ nm~ $w_2 = w_3 = 10$ nm for CuInP$_2$S$_6$. Compare the big difference in shifts $x_1 - x_2 = -215$ nm, $x_3 - x_4 = -185$ nm for CuInP$_2$S$_6$ with small and the same difference $x_1 - x_2 = x_3 - x_4 = -5$ nm for CuInP$_2$Se$_6$. Exactly the difference determines the cross-over from the "humps" at bright walls to zero values at dark walls. Note that the alternating signs of $u_i$ are all the same for both sulfide and selenide compounds. These trends are in a qualitative agreement with the tanh-like fitting by Eqs.(11) of the FEM solution of Eqs.(10). Actually for $\epsilon > 0$, $p_s$ is smaller than $a_s$, and $L_p$ is higher than $L_a$. The situation is opposite for $\epsilon < 0$, when $p_s > a_s$ and $L_p < L_a$.

Based on the domain wall widths, extracted from **Fig. 5a**, FEM results shown in **Fig. 4d**. Intuitively the result is clear, because **Fig. 4c** contains wider regions of brightness, and the existence of bright domains in a ferrielectric CuInP$_2$Se$_6$ is something "mixed" between FE and AFE state.

## VI. CONCLUSION

The main result of this work is the explanation of the unusual domain wall structure in FE, AFE and mixed AFE-FE phases of Cu-based layered ferrielectric CuInP$_2$(S$_{1-y}$Se$_y$)$_6$. In accordance with LGD-FSM approach, proposed by us earlier [32, 33], the spatial gradient of the local Curie temperature, can lead to the coexistence of the FE, AFE and spatially modulated ferrielectric FE-AFE phases, as well as to the suppressed or enhanced local electromechanical contrast at the boundaries between the phases. Note, that the aforementioned behaviors originated from the system tendency to minimize LGD-FSM free energy under the certain conditions imposed on the control parameters (Curie temperatures variation $\epsilon$ and the coupling strength $\chi$ between FE and AFE orders).

Since "dark" or "bright" boundaries have been recently observed by PFM and optical spectroscopies experiments for CuInP$_2$(S$_{1-y}$Se$_y$)$_6$ (where *y*=0 and *y*=1), our theoretical results being in quantitative agreement with the experiments [31], provides insight to the origin of unusual domain boundaries in FE-AFE layered ferrielectrics.

**Acknowledgements.** Authors gratefully acknowledges Dr. J. A. Brehm for useful discussions. Authors acknowledge Lei Tao, Andrew O'Hara and Sokrates Pantelides for sharing calculated structures of antiferroelectric CuInP$_2$Se$_6$. This material is based upon work (S.V.K, P.M.) supported by the Division of Materials Science and Engineering, Office of Science, Office of Basic Energy Sciences, U.S. Department of Energy, and performed in the Center for Nanophase Materials Sciences, supported by the Division of Scientific User Facilities. A portion of FEM was conducted at the Center for Nanophase Materials Sciences, which is a DOE Office of Science User Facility (CNMS Proposal ID: 257). A.N.M work is supported by the National Academy of Sciences of Ukraine (the Target Program of Basic Research of the National Academy of Sciences of Ukraine "Prospective basic research and innovative development of nanomaterials and nanotechnologies for 2020 - 2024", Project № 1/20-H, state registration number: 0120U102306) and has received funding from the National Research Foundation of Ukraine (Grant application 2020.02/0027).

**Authors' contribution.** A.N.M., P.M. and S.V.K. generated the research idea and propose the theoretical model. A.N.M. derived analytical results, interpreted numerical results, obtained by E.A.E, compare with experiment, performed by K.K. and P.M., and wrote the manuscript draft. S.V.K., Y.M.V. and P.M. worked on the results discussion and manuscript improvement.

## APPENDIX A

The free energy (10) as a function of polar and antipolar order parameters, $p$ and $a$, is:

$$g_{LGD} = -(1-\epsilon)\frac{p^2}{2} - (1+\epsilon)\frac{a^2}{2} + \frac{p^4+a^4}{4} + \frac{\chi}{2}p^2 a^2 + \frac{g}{2}\left[\left(\frac{dp}{dx}\right)^2 + \left(\frac{da}{dx}\right)^2\right], \quad (A.1)$$

The properties of bulk homogeneous system with $dp/dx = 0$ and $da/dx = 0$ are considered below. Equation of state are

$$-(1-\epsilon)\mathrm{p} + p^3 + \chi\, p\, a^2 = 0, \quad (A.2a)$$
$$-(1+\epsilon)\mathrm{a} + a^3 + \chi\, p^2 a = 0, \quad (A.2b)$$

The solution of (A.2a)-(A.2b) determine the phase of the system. Only the stable solutions are physically permissible, hence the matrix of the second derivatives of the free energy (A.1) with respect to order parameters $a$ and $p$ should be positively defined:

$$\left\| \frac{\partial^2 g_{bulk}}{\partial p\, \partial a} \right\| = \begin{pmatrix} -(1-\epsilon) + 3p^2 + \chi a^2 & 2\chi p a \\ 2\chi p a & -(1+\epsilon) + 3a^2 + \chi p^2 \end{pmatrix} \quad (A.3)$$

Four spatially-homogeneous phases could exist for the system with free energy (A.1), namely:

1) Para-phase **P-phase** with the order parameters and energy density

$$p = 0, \quad a = 0 \quad (A.4a)$$

$$g_{bulk} = 0 \quad (A.4b)$$

Stability matrix (A.3) for this case is

$$\left\| \frac{\partial^2 g_{bulk}}{\partial p\, \partial a} \right\| = \begin{pmatrix} -(1-\epsilon) & 0 \\ 0 & -(1+\epsilon) \end{pmatrix} \quad (A.4c)$$

Hence **P-phase** is unstable (for the chosen form of the free energy (A.1) coefficients)

2) Tetragonal **$T_p$-phase** (see **Figs. 2d-e**) with the order parameters and energy density:

$$p = \pm\sqrt{1-\epsilon}, \quad a = 0 \quad (A.5a)$$

$$g_{bulk} = -\frac{(1-\epsilon)^2}{4} \quad (A.5b)$$

Stability matrix (A.3) for this case is

$$\left\| \frac{\partial^2 g_{bulk}}{\partial p\, \partial a} \right\| = \begin{pmatrix} 2(1-\epsilon) & 0 \\ 0 & -(1+\epsilon) + \chi(1-\epsilon) \end{pmatrix} \quad (A.5c)$$

Hence, **$T_p$-phase** is stable at

$$\chi > \chi_{cr}^p(\epsilon) \stackrel{def}{=} \frac{1+\epsilon}{1-\epsilon} \quad \text{and} \quad \epsilon < 1 \quad (A.5d)$$

3) Tetragonal **$T_a$-phase** (see **Figs. 2d-e**) with the order parameters and energy density:

$$p = 0, \quad a = \pm\sqrt{1+\epsilon} \quad (A.6a)$$

$$g_{bulk} = -\frac{(1+\epsilon)^2}{4} \quad (A.6b)$$

Stability matrix (A.3) for this case is

$$\left\| \frac{\partial^2 g_{bulk}}{\partial p\, \partial a} \right\| = \begin{pmatrix} -(1-\epsilon) + \chi(1+\epsilon) & 0 \\ 0 & 2(1+\epsilon) \end{pmatrix} \quad (A.6c)$$

Hence, **$T_a$-phase** is stable at

$$\chi > \chi_{cr}^a(\epsilon) \stackrel{def}{=} \frac{1-\epsilon}{1+\epsilon} \quad \text{and} \quad \epsilon > -1 \quad (A.6d)$$

4) Orthorhombic **O-phase** (see **Figs. 2a-c**) with the order parameters and energy density:

$$p = \pm\sqrt{\frac{1-\epsilon-\chi(1+\epsilon)}{1-\chi^2}}, \quad a = \pm\sqrt{\frac{1+\epsilon-\chi(1-\epsilon)}{1-\chi^2}} \quad (A.7a)$$

$$g_{LGD} = -\frac{1+\epsilon^2-\chi(1-\epsilon^2)}{2(1-\chi^2)} \quad (A.7b)$$

Stability matrix (A.3) for this case is

$$\left\|\frac{\partial^2 g_{bulk}}{\partial p\, \partial a}\right\| = \begin{pmatrix} 2p^2 & 2\chi\, p\, a \\ 2\chi\, p\, a & 2a^2 \end{pmatrix} \quad (A.7c)$$

One could see from (A.7c) that O-phase is stable at $p^2 > 0, a^2 > 0$ and $\chi^2 < 1$, which gives the following conditions:

$$-\frac{1-\chi}{1+\chi} < \epsilon < \frac{1-\chi}{1+\chi} \quad \text{and} \quad \chi^2 < 1 \quad (A.7d)$$

$T_p$-$T_a$ equilibrium is achieved under the condition $\epsilon = 0$ at $\chi > 1$; O-$T_a$ equilibrium is achieved under the condition $\chi = \chi_{cr}^a(\epsilon)$ at $\chi^2 < 1$, O-$T_p$ equilibrium is achieved under the condition $\chi = \chi_{cr}^p(\epsilon)$ at $\chi^2 < 1$.